# Horizontal subduction zones, convergence velocity and the building of the Andes.

J. Martinod[1], L. Husson[2,3], P. Roperch[1,2], B. Guillaume[4] and N. Espurt[5]

(1) Corresponding author ; LMTG, UMR CNRS-IRD-Université de Toulouse, 14 avenue Edouard Belin, 31400 Toulouse, France (martinod@lmtg.obs-mip.fr). Tel (+33) 5 61 33 26 52 ; fax (+33) 5 61 33 25 60

(2) Géosciences Rennes, UMR-CNRS 6118, Université Rennes 1, 35042 Rennes cedex, France

(3) Laboratoire de Planétologie et Géodynamique, UMR CNRS 6112, université de Nantes, BP 92208, 44322 Nantes cedex, France

(4) Università Roma TRE, Dipartimento Scienze Geologiche, Largo S. L. Murialdo 1, 00146 Roma, Italy

(5) CEREGE, UMR-CNRS, Aix-Marseille Université, Europôle Méditerranéen de l'Arbois, BP 80, 13545 Aix en Provence cedex 04, France

*Abstract :*

We discuss the relationships between Andean shortening, plate velocities at the trench, and slab geometry beneath South America. Although some correlation exists between the convergence velocity and the westward motion of South America on the one hand, and the shortening of the continental plate on the other hand, plate kinematics neither gives a satisfactory explanation to the Andean segmentation in general, nor explains the development of the Bolivian orocline in Paleogene times. We discuss the Cenozoic history of horizontal slab segments below South America, arguing that they result from the subduction of oceanic plateaus whose effect is to switch the buoyancy of the young subducting plate to positive. We argue that the existence of horizontal slab segments, below the Central Andes during Eocene-Oligocene times, and below Peru and North-Central Chile since Pliocene, resulted (1) in the shortening of the continental plate interiors at a large distance from the trench, (2) in stronger interplate coupling and ultimately, (3) in a decrease of the trenchward velocity of the oceanic plate. Present-day horizontal slab segments may thus explain the diminution of the convergence velocity between the Nazca and South American plates since Late Miocene.

## Introduction

The relationship between the uplift of the Andean Cordillera and the convergence velocity at the trench has been debated for several decades. In the one hand, some authors note that Andean building has largely been controlled by plate kinematics, either by the convergence velocity between the oceanic plate and overriding continent (e.g., Pardo-Casas and Molnar, 1987; see below), or by the absolute westward motion of South America (e.g., Silver et al., 1998). In the other hand, others remark that the Andean relief results in stresses that may affect the trenchward velocity of the subducting plate (e.g., Iaffaldano et al., 2006). Both hypotheses are supported by temporal correlations between Andean tectonics, Andean uplift and plate kinematics around South America. This paper focuses on an additional phenomenon, the appearance of horizontal slab segments. Indeed, we argue that slabs that subduct horizontally largely modify the overriding plate tectonic regime that in turn feedbacks on the dynamics of oceanic subduction. If flat-slab subduction coevally changes the rate of uplift of the Andes and the subduction velocity, it creates temporal correlations between them that may be misleading.

In the following, we review the data that plead for an influence of the convergence velocity on Andes building. We also review the arguments conversely suggesting that Andean growth influenced the trenchward oceanic plate velocity. Then, we infer the Cenozoic history of horizontal subduction zones beneath South America, looking at the geometry and the evolution of the volcanic arc. We show that, in fact, flat slabs may have largely influenced both the subduction velocity and the Andean orogenesis.

**Impact of plate kinematics on Andean shortening**

Authors early noted that the growth of the Andes is recent compared to the long-lived oceanic subduction along the western margin of South America, that has been active at least since the beginning of the Jurassic (Coira et al., 1982; see also e.g. Pankhurst et al., 2000, for Patagonia; Oliveros et al., 2006; Mamani et al., 2010, for the Central Andes). Conversely, crustal thickening in the overriding plate only started in the Upper Cretaceous whereas Jurassic and Lower Cretaceous times have essentially been marked by arc and back-arc extension (e.g., Mpodozis and Ramos, 1989; Jaillard and Soler, 1996). Jaillard and Soler (1996), following several authors, observe that tectonic shortening occurred during three major discrete pulses in the Upper Cretaceous, Eocene, and Neogene (Figure 1). Other periods during the Cenozoic, in contrast, are marked by moderate shortening and even locally by extension.

(1) The Upper Cretaceous *Peruvian* shortening event has been widely observed from Ecuador (Jaillard et al., 1996; Jaillard et al., 2005) to Peru (Vicente, 1989; Jaillard, 1994), Bolivia (Sempere, 1994), and Southern Andes (e.g., Cobbold and Rossello (2003) for the Neuquen basin; Diraison et al. (2000), Kraemer (2002), Ramos (2002), for Central and Southern Patagonia).

(2) A second major Eocene shortening event occurred in the Central Andes. It corresponds to the *Incaic* phase of Steinmann (1929) (Jaillard et al., 1996; Lamb et al., 1997). Continental shortening is also observed in the Southern Andes at that time (Cobbold and Rossello, 2003). Oligocene, in contrast, is marked by a decrease of the shortening rate in the Western Cordillera of Central Andes. Oligocene times also correspond to a diminution of the shortening rate in the Southern Andes, some authors arguing that continental extension occurs at that time between at least 20°S and 44°S (e.g., Pananont et al., 2004; Zapata and Folguera, 2005; Jordan et al., 2007; see also Charrier et al., 2005, for a review), while others doubt that significant extension really occurred (Cobbold and Rossello, 2003; Arriagada et al., 2006). It is remarkable that this period also corresponds to the onset of shortening in the Eastern Cordillera of the Central Andes (e.g., Lamb and Hoke, 1997; Roperch et al., 2006; Oncken et al., 2006).

(3) The last widely reported major shortening episode within the Andean belt is the Neogene *Quechua* phase, observed along the entire Cordillera, from Colombia to Patagonia. Shortening is still active; in the Central Andes, surface shortening is accommodated in the Subandean Zone (e.g., Lamb, 2000).

The existence of discrete periods of Andean shortening, within a long-term process of oceanic subduction, has early been observed and discussed. For Uyeda and Kanamori (1979), Cross and Pilger (1982) or Pardo-Casas and Molnar (1987), shortening of the overriding plate is produced by the rapid convergence between the oceanic and continental plates. As a matter of fact, the three major shortening episodes described above correspond to periods of rapid trench-perpendicular convergence between the subducting plate and South America, while convergence has been slower and/or more oblique during both the Maastrichtian-Early Paleocene period and the Oligocene (Pardo-Casas and Molnar, 1987; Soler and Bonhomme, 1990; Sdrolias and Muller, 2006) (Figure 1).

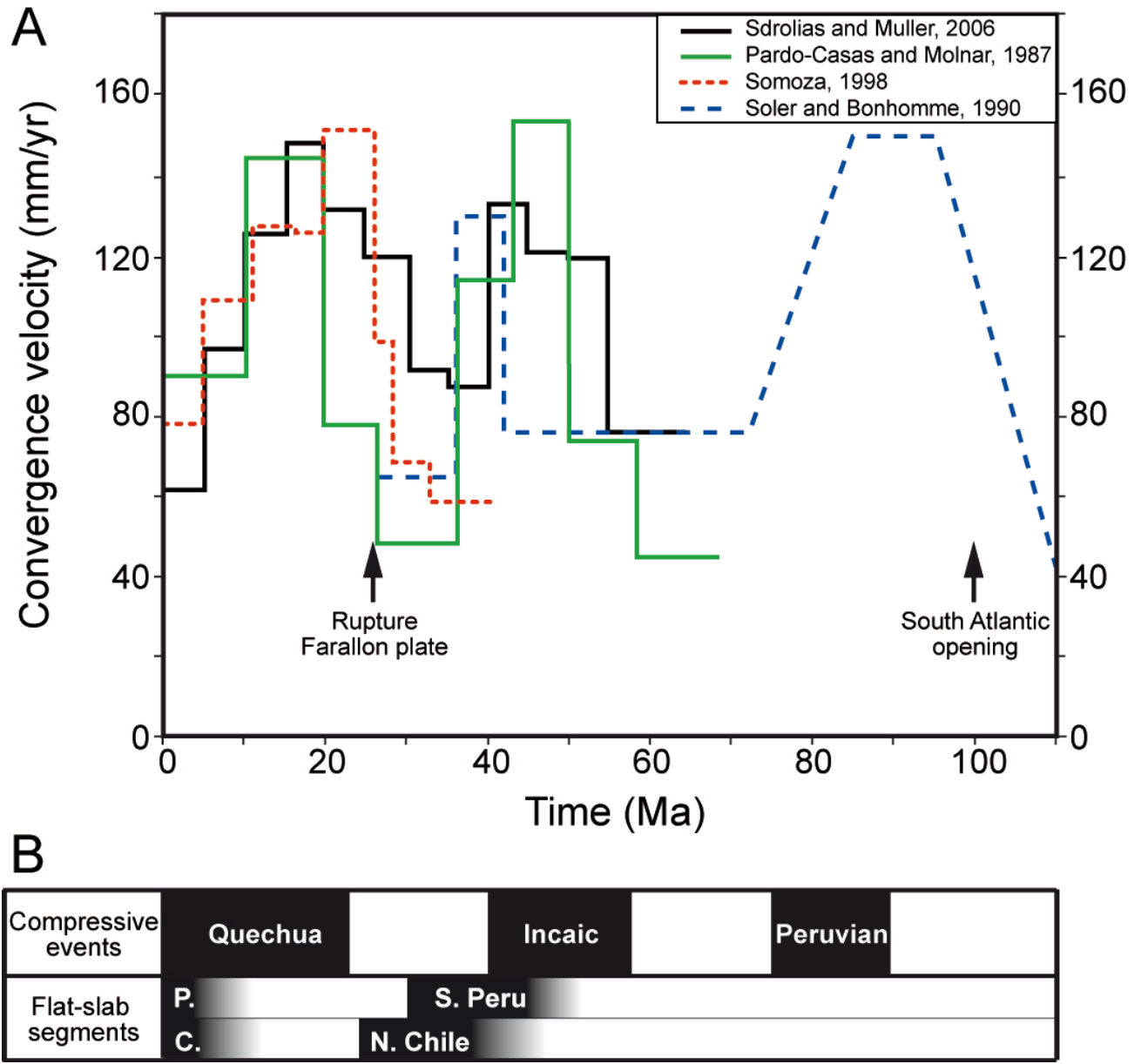

**Figure 1: (A)** Convergence velocity (mm/yr) between the Farallon-Nazca oceanic plate and South America, at the latitude of the Central Andes, from Pardo-Casas and Molnar (1987), Bonhomme and Soler (1990), Somoza (1998), and Sdrolias and Muller (2006). 23 Ma-old break-up of the Farallon plate after Lonsdale (2005). **(B)** Major episodes of continental shortening after Steinman (1929) and Mégard et al. (1984), and periods of horizontal subduction beneath parts of South America. P.: Peru; C.: Chile; Flat-slab subductions correlate with slower convergence velocity at trench.

The absolute westward velocity of the South American plate, on the other hand, may also have controlled the long-term shortening of the active margin from the Upper Cretaceous (Frutos, 1981; Jarrard, 1986). The westward acceleration of South America started with the opening of the equatorial segment of the Atlantic Ocean ~100 Ma ago, which approximately corresponds to the beginning of shortening in the continental plate. Silver et al. (1998) attribute the Neogene episode of the Andean orogeny to the increase of the westward drift rate of the South American plate, induced by the collision of Africa onto Eurasia that opposes further northeastward motion of the African plate. Sdrolias and Muller (2006), however, do not observe any major increase of the South American plate absolute westward velocity during the Neogene. In fact, the onset of Neogene shortening is also coeval to the break-up of the Farallon plate into the Nazca and Cocos plates, which resulted in a rapid trench-perpendicular motion of the subducting plate beneath South America (Pardo Casas and Molnar, 1987; Somoza, 1998).

The above-mentioned data suggest some correlation between Andean shortening and both the trench-perpendicular convergence velocity and the absolute westward motion of South America (Figure 1). This correlation, however, is not perfect and for instance, Charrier et al. (2002; 2005) observe that the beginning of Neogene shortening in the Southern Andes does not occur precisely at the same moment at different latitudes. The onset of shortening in the Cordillera of Central Chile is ~25 Ma old at 31.5 °S, 21 Ma old at 33°S, and the inversion of the Oligocene extensional basin only began 16 Ma ago in the Cachapoal basin (34.5°S) (Charrier et al., 2005). Furthermore, in spite of the decreasing convergence rate to its slowest value in Neogene times (Pardo Casas and Molnar, 1987; Somoza, 1998) and maybe for the last 60 Ma (Sdrolias and Muller, 2006), present-day shortening rates remain as high as 7 to 17 mm/a in the Central Andes (Lamb, 2000). Thus, it is clear that plate kinematics alone cannot explain the growth of the Andes, and particularly its latitudinal segmentation.

### Andean segmentation

The Andes show remarkable latitudinal variations in their maximum mean elevation, width, and crustal shortening, despite Cenozoic plate velocities at the trench have been roughly similar in the Central and Southern Andes, north of the Nazca-Antarctica-South America triple junction. Large uncertainties remain on the trench-perpendicular crustal shortening accommodated in the Bolivian orocline, which has been estimated between 200 and 500 km (Baby et al., 1997; Kley and Monaldi, 1998; McQuarrie, 2002; Hindle et al., 2005; Arriagada et al., 2008). Shortening in the Central Andes is in any case much larger than in its Southern and Northern counterparts. South of 36°S, for instance, the total crustal shortening may not exceed 50 km (e.g., Kley and Monaldi, 1998).

Different causes have been invoked to explain the Andean segmentation and the particularly large development of Central Andes if compared to other segments of the belt. Lamb and Davis (2003) propose that the hyper-arid climate of the Pacific side of the Central Andes resulted in sediment starvation within the oceanic trench, which, in turn, would have increased interplate friction and triggered continental shortening. Another explanation has been proposed by Russo and Silver (1994) and reformulated by Schellart et al. (2007) to explain the larger Central Andean shortening. These authors observe that the Central Andes are located far from the lateral edges of the Nazca subducting plate. The slab retreat that accommodates the westward drift of South America would be more difficult in the Central Andes, because the underlying asthenosphere cannot easily laterally bypass the slab in a toroidal flow from the Pacific mantle reservoir to the Atlantic one. Slower slab retreat below the Central Andes would explain the larger Cenozoic continental shortening in that part of the Cordillera provided that South America has a wholesale westward motion.

The Bolivian orocline formed as a result of the larger shortening in Central Andes (e.g., Isacks, 1988; Arriagada et al., 2008). Paleomagnetic data from the Central Andes forearc reveal, however, that this orocline essentially appeared in Eocene-Oligocene times (Roperch et al., 2006; Arriagada et al., 2008). Although Neogene rotations have been described in the backarc of the Central Andes (e.g., Macfadden et al., 1995; Rousse et al., 2002), rocks younger than 25 Ma do not show evidence for rotations in the forearc (Roperch et al., 2006). It indicates that the bending of the orocline essentially stopped 25 Ma ago, and that the Neogene rotations that are observed within the backarc domain result from local block rotations. Thus, the present-day geometry of the Nazca subduction zone shall not be invoked to explain the orocline. One should rather consider the geometry of the earlier, Eocene-Oligocene Farallon slab, and it is not clear that the distance from the lateral edges of the Farallon slab may similarly explain the stronger shortening in Central Andes at that time. Instead, we propose below that the Paleogene formation of the orocline may have been largely controlled by the existence of a horizontal slab segment below the Central Andes.

**Impact of Andean elevation on plate kinematics**

If in the one hand, plate kinematics affects the tectonics of the active margin, on the other hand, the Andean relief results in compressive stresses on the neighboring lowland areas (e.g., Dalmayrac and Molnar, 1981; Froidevaux and Isacks, 1984), that may in turn decrease the trenchward velocity of the subducting plate (Iaffaldano et al., 2006; Meade and Conrad, 2008). Furthermore, it has been proposed that Andean forcing on the volume of the Pacific upper mantle may even cause the present-day westward drift, in the hot spot reference frame, of the Pacific Basin (Husson et al., 2008).

Most authors now agree that the Altiplano acquired most of its present-day ~4000 m high topography during the last 10 Ma. This has been proposed from analyses of (1) paleovegetation (Gregory-Wodzicki, 2000), (2) oxygen isotopic composition of carbonates deposited in the Altiplano (Garzione et al., 2006), (3) tectonic regime showing a cessation of shortening within the Altiplano and its transfer towards the Subandean zones (Lamb and Hoke, 1997), (4) geomorphological evolution of the Chilean and Peruvian Precordillera, in which the Late Miocene appearance of deep canyons resulted from the gentle tilting of that part of the fore-arc region (Lamb et al., 1997; Lamb and Hoke, 1997; Farias et al., 2005; Riquelme et al., 2007; Thouret et al., 2007). However, the rapid uplift of the Central Andes since Late Miocene is contested by Barnes and Ehlers (2009). They suggest instead that the shortening of the Andes and subsequent exhumation are much older and they propose that Andean uplift may have occurred progressively for 25 Ma. Parts of the Central Andes, indeed, may have achieved an elevation as high as 2000-2500 m in Early Miocene times (Riquelme et al., 2003; Picard et al., 2008).

Garzione et al. (2006) and Iaffaldano et al. (2006) propose that the growth of the Andean relief is responsible for the deceleration of the Nazca subduction rates that occurred since 10 Ma (Somoza, 1998). Garzione et al. (2006) argue that only the removal of dense eclogitic lower crust and/or mantle lithosphere could generate the very rapid uplift of the Central Andes they observe. This lithospheric unloading below the chain would also explain the transfer of compressive deformation towards the Subandean zones, and the simultaneous overfilling of the Amazonian foreland basin (Roddaz et al., 2006). Then, two distinct periods would mark the evolution of the central part of the Cordillera. A constructive period starting in the Upper Cretaceous and finishing in the Miocene is essentially marked by the thickening of the continental lithosphere. Following the Late Miocene uplift, the Central Andes would have shifted to a mature plateau stage, during which the elevation of the belt can no longer increase. The Andean relief would in turn transmit compression on the lowlands of the Sub Andean zone, and would slow down the subduction of the Nazca plate.

In the following however, we argue that the geometry of the Nazca slab itself may better explain the decreasing trenchward velocity of the oceanic plate.

**Horizontal subduction underneath the Andes and subduction of buoyant slab segments**

Subduction zones in which the descending plate sinks in the asthenosphere with a small dip angle are relatively common (see e.g., Lallemand et al., 2005). In contrast, there are few slab segments whose geometry is that of the present-day Andean flat-slab subduction zones. Seismological data gives a satisfactory description of the current geometry of the subducting plate below the Andes (Barazangi and Isacks, 1979; Cahill and Isacks, 1992; Pardo et al., 2002; Gilbert et al., 2006). Two major flat-slab segments are imaged below South America (Figure 2), the largest one being located below Central and Northern Peru between 3°S and 15°S. The second flat-slab segment is located between 28° and 33°S, beneath Central Chile and Argentina. In both segments, the subducting plate descends with a maximum dip of 30° from the trench to a depth of approximately 100-120 km, and then flattens underneath the overriding lithosphere for several hundreds of kilometers before sinking into the upper mantle asthenosphere. The asthenospheric corner is repelled as far as 700 km and 600 km away from the trench, in Peru and Argentina, respectively.

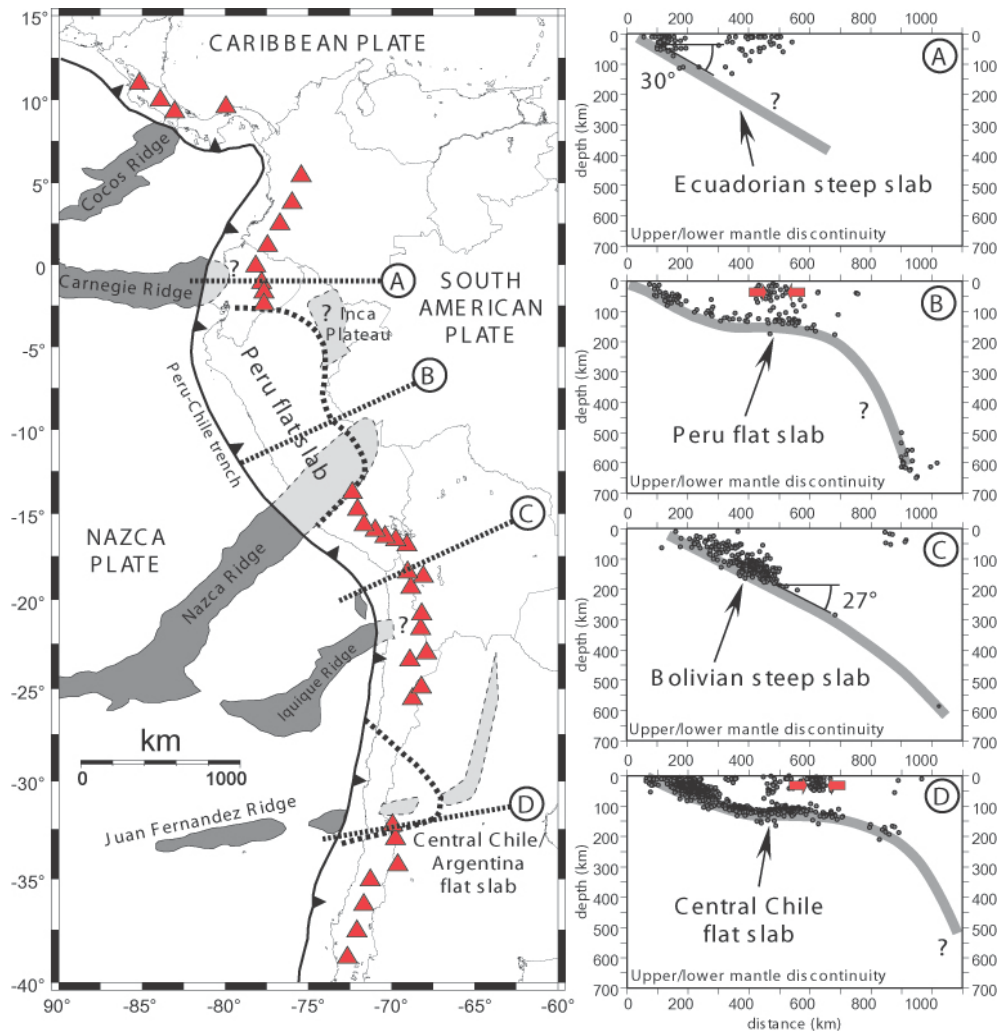

**Figure 2: Present-day configuration of the Cordillera, after Espurt et al. (2008). Triangles mark active volcanoes. Flat-slab segments and oceanic ridges have been reported.**

The subducting Nazca plate is younger than 50 Ma along the entire length of the trench, which suggests that its negative buoyancy is at best low (e.g., Cloos, 1993). Therefore, slight density perturbations may easily affect its downward route into the mantle. Analogue and numerical models suggest that horizontal slabs can result from the forced subduction of buoyant slab segments below an advancing overriding plate (Van Hunen et al., 2002; Martinod et al., 2005; Espurt et al., 2008). Because the thermal structure of the young Nazca slab minimizes its negative buoyancy, its Stokes velocity is low, i.e. its vertical descent is slow. If the Atlantic Ocean is to expel the South American plate to the West at a faster rate than the vertical velocity of the slab, its dip decreases accordingly and the slab flattens underneath the overriding plate.

In addition, departure from the negative slab buoyancy that drives the slab to depth may be achieved by slight modification of its density structure (e.g., Davila et al., 2010). In particular, authors early noted that the present-day flat-slab segments beneath South America may be explained by the subduction of buoyant ridges (e.g. Sacks, 1983; Henderson et al., 1984; Pilger, 1984). The southern limit of the Peruvian flat-slab along the trench indeed coincides with the Nazca ridge. Gutscher et al. (2000) proposed that the Northern part of the Peruvian slab is held flat by the subduction of an oceanic plateau ("Inca Plateau"), which would explain why the trench-parallel extension of this horizontal slab segment exceeds 1000 km. The southern boundary of the Chilean-Argentinean flat-slab segment also precisely coincides with the subduction of a passive ridge coming from the Juan Fernandez hot spot (Yañez et al., 2001). The Juan Fernandez ridge is marked in the South-East Pacific Ocean bathymetry by a moderate and discontinuous anomaly four times smaller than the Nazca Ridge. The corresponding thickness of the oceanic crust is not sufficient to make the oceanic plate buoyant (Kopp et al., 2004; Martinod et al., 2005), but the present-day topography of the Pacific Ocean only gives information on the last 10 Ma hot spot magmatic activity. If slab buoyancy promotes slab flattening, the volume of magma that resulted from the Juan Fernandez hot spot activity in the past must have been larger (Espurt et al., 2008), although other phenomena such as upper mantle hydration may have further decreased the subducting plate density (Kopp et al., 2004).

The occurrence of horizontal slab segments beneath South America can be inferred by looking at the evolution of the arc volcanism in South America. Indeed, authors early noted that present-day horizontal slabs beneath South America are marked by an absence of any volcanic activity (e.g., Pilger, 1981; Nur and Ben-Avraham, 1981; McGeary et al., 1985). The spatial distribution of arc volcanism broadens when the horizontal subduction is forming because the volcanic activity progressively migrates away from the trench. It eventually switches off when the asthenospheric wedge above the slab gets pinched enough to cool the overriding plate and starve the continent from its magmatic contribution (Kay et al., 1988; James and Sacks, 1999; Ramos et al., 2002; Kay and Mpodozis, 2002; Bourdon et al., 2003). Therefore, there is a delay between the onset of slab flattening and volcanic quiescence, which only marks the ultimate development of slab flattening. Espurt et al. (2008) noted that arc volcanism ceased ~7 Ma after the arrival at the trench of buoyant oceanic lithosphere both in Peru and Chile-Argentina. This timing is further supported by the results of both their analogue models and the numerical models of Hassani et al. (1997). Espurt et al. (2008) suggest that slab flattening that results from the subduction of the Iquique and Carnegie ridges may be active yet not over, because the subduction of these ridges below South America is too recent (1-2 Ma, Lonsdale and Klitgord, 1978; Rosenbaum et al., 2005). Active slab flattening resulting from the subduction of the Carnegie ridge is indeed supported by the analysis of the geographical and geochemical characteristics of Quaternary volcanism in Northern Ecuador (Bourdon et al., 2003). The impact of the Iquique Ridge on the present-day process of subduction is not clear, probably because this ridge is much smaller than the Carnegie Ridge.

The analysis of arc volcanism reveals that the two major flat-slab segments below South America were present at least during Pliocene times (Figure 1). In central Peru, indeed, arc volcanism became inactive ~4 Ma ago (Soler and Bonhomme, 1990; Rosenbaum et al., 2005). In Central Chile-Northwestern Argentina, arc volcanism has been restricted to the Sierras Pampeanas for 5 Ma, and the end of volcanic activity is ~2 Ma-old (Kay and Gordillo, 1994; Ramos et al., 2002). Conversely, during the Late Miocene, arc volcanism was active along the entire Cordillera (e.g., James and Sacks, 1999, for the Central Andes; Kay and Mpodozis, 2002, for the South Central Andes), revealing the absence of any fully developed horizontal subduction at that time below South America. Several reasons may explain why the subduction of the Juan Fernandez ridge was not accompanied by any horizontal slab segment in the Late Miocene. Either the volume of the ridge that had subducted at that time was too small to make the slab buoyant, or horizontal slab segments had not enough time to appear because of the rapid Miocene southward migration of the ridge beneath South America. Since ~11 Ma, the ridge subducted approximately at the same latitude beneath Chile (Yañez et al., 2001), which may explain why a horizontal slab segment appeared in the Pliocene.

Another major horizontal slab segment seems to have existed below the Central Andes in the Paleogene. As a matter of fact, James and Sacks (1999) underline that in Southern Peru and Northern Chile, the volcanic activity almost ceased during approximately 20 Ma in the Tertiary. They argue that the characteristics of the volcanism in this part of the Andes resulted from changes in the geometry of the slab. Mamani et al. (2010) show that the magmatic arc, whose position in the Central Andes had remained stable from the beginning of Upper Cretaceous, switched to the interior of the continent during the Late Eocene. In Southern Peru, the anomalous position of the magmatic arc lasted between about 45 and 30 Ma (Mamani et al., 2010). In Northern Chile, volcanism ceased between 38 Ma and the end of the Oligocene (Figure 1, Hammerschmidt et al., 1992; Soler and Jimenez, 1993). Early Miocene corresponds to the onset of widespread mafic volcanism in the Altiplano (Lamb and Hoke, 1997) and to the beginning of ignimbrite deposition in the Chilean forearc (Wörner et al., 2000; Farias et al., 2005). The reactivation of volcanism occurred very suddenly (Wörner et al., 2000) and almost simultaneously in both the Altiplano and the Western Cordillera. The renewal of volcanism in northern Chile, 24 Ma ago, can be interpreted as a result of the end of the flat-slab episode there.

If the subduction of buoyant slab segments explains the present-day geometry of the Nazca slab, it may also have triggered slab flattening in the Paleogene underneath the Bolivian orocline. Plate reconstructions suggest, indeed, that if an oceanic plateau formed above the Juan Fernandez hot spot during the Upper Cretaceous and was transported above the Farallon plate, it may have reached the Southern Peruvian trench during the Eocene (Figure 3). To reconstruct the trail of the Juan Fernandez hot spot and its subduction below the South American plate, it is necessary (1) to assume that the present volcanic center in the Juan Fernandez archipelago corresponds to an active hot spot and (2) to know the absolute motion of the Nazca-Farallon plate through time. Whether hot spots are fixed and can be used to define an absolute reference frame is a matter of major controversies (e.g., Anderson, 2005; Tarduno et al., 2003; 2009). For instance, Molnar and Stock (1987) first showed that significant discrepancies exist between the Indo-Atlantic and the Pacific hot spot reference frame. We calculated the trails of the Juan Fernandez hot spot assuming three different configurations. The first configuration considers the plate kinematic parameters used by the Earthbyte group and Muller et al. (2008), with the absolute motion of the Nazca/Farallon plate tied to the Indo-Atlantic moving hot spot reference frame of O'Neill et al. (2005). The second configuration uses the absolute motion of the Pacific plate determined by Wessel and Kroenke (2008), the Nazca/Farallon plate being linked to this reference frame while the absolute motion of South America is calculated within the Indo-Atlantic reference frame. In that case we assume that both hot spot systems are fixed, that the plate circuit around Antarctica is not reliable and/or that West and East Antarctica followed different circuits for 100 Ma. For the last configuration, the absolute motion defined by Gordon and Jurdy (1986) for Nazca/Farallon and South America is used as in Yañez et al. (2001).

Differences in the absolute motion of the Nazca/Farallon plate lead to three discrepant trails (Figure 3). The three reconstructions, however, predict that segments of the Juan Fernandez ridge subducted below the Central Andes of Southern Peru during the Late Eocene-Oligocene. The ridge segment that subducted during the Oligocene beneath South America was ~75 Ma-old (Figure 3). It is hazardous to reconstruct the motion of the ridge segment that formed earlier, e.g. between 100 and 75 Ma. Plate reconstructions suggest it may have subducted either below Central or Southern Peru during the Eocene if the ridge emplaced above the Farallon plate (Figure 3). Note however that in the plate reconstructions of Sdrolias and Müller (2006), Müller et al. (2008) and Torsvik et al. (2010), the spreading ridge separating the Phoenix (or Aluk) plate from the Farallon plate was located close to the Juan Fernandez hot spot in Late Cretaceous times (Figure 4). The position of the ridge separating the Phoenix and Farallon plates is poorly constrained by magnetic anomalies (e.g., Breitsprecher and Thorkelson, 2009) since most of the ridge has been subducted. Thus, any plateau formed above this hot spot before 80 Ma may also have moved to the South along with the Phoenix plate and may not have interacted with the Central Andes.

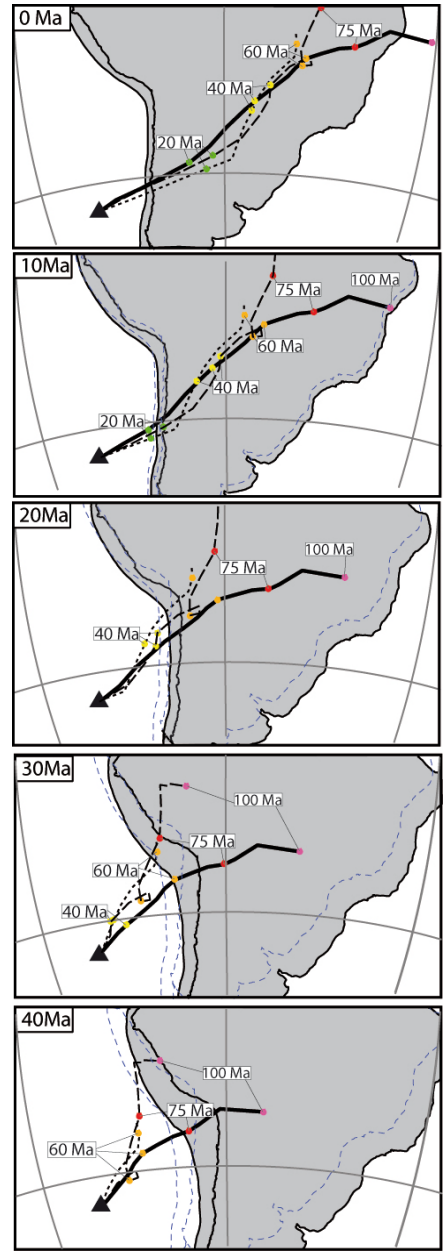

**Figure 3: Reconstruction of the trails of the Juan Fernandez hot spot for the last 40 Ma. Gray shaded areas correspond to South America. The western border of South America is underformed through time to take into account the oroclinal bending associated with shortening proposed by Arriagada et al. (2008). Trails have been obtained supposing that the Juan Fernandez hot spot has been located beneath the Farallon-Nazca plate during the last 100 Ma, and considering either (1) plate kinematics parameters for 100 Ma in the Indo-Atlantic moving hot spot reference frame (Muller et al., 2008) (dashed line); (2) the absolute motion for 100 Ma of the Farallon-Nazca plate in the Pacific reference frame of Wessel and Kroenke (2008) (solid line); (3) Gordon and Jurdy (1986) absolute plate motion for 64 Ma (dotted line and dotted contours for the South American continent). See text for further details.**

The strongest available clue on the history of ridge subduction beneath South America is the registry of magmatic activity. Until Cenozoic times the locus of magmatism in Peru and Chile exhibited relatively straightforward behavior, the trenchward magmatic front slowly migrating eastward (Soler and Bonhomme, 1990; James and Sacks, 1999). In Central Peru, the magmatic arc, whose position had moved eastward no more than 50 km between 100 and 40 Ma, widened to the east ~40 Ma ago (Soler and Bonhomme, 1990; Bissig et al., 2008). Arc widening has been attributed to a diminution of the dip of the slab. Magmatism, however, remained active there, suggesting that the Paleogene horizontal slab segment was restricted to Southern Peru and Northern Chile (James and Sacks, 1999). The Eocene diminution of the slab dip beneath Central Peru may have resulted from the Eocene trenchward acceleration of South America (Soler and Bonhomme, 1990). Indeed, statistical analyses of modern subduction zones as well as models confirm that slab dip diminishes when the trenchward absolute motion of the overriding plate increases (Lallemand et al., 2005; Heuret et al., 2007). Thus, we do not have any piece of evidence that segments of the Juan Fernandez Ridge subducted below Central Peru in the Paleogene. If so, these segments were not large enough to balance the negative buoyancy of the oceanic plate and to trigger horizontal subduction. In contrast, the appearance of a major volcanic plateau above the Juan Fernandez hot spot during the Campanian, migrating northward above the Farallon plate and finally subducting beneath the Central Andes during the Late Eocene and Oligocene, is a plausible explanation for the appearance of the above described Paleogene horizontal subduction (Figure 4).

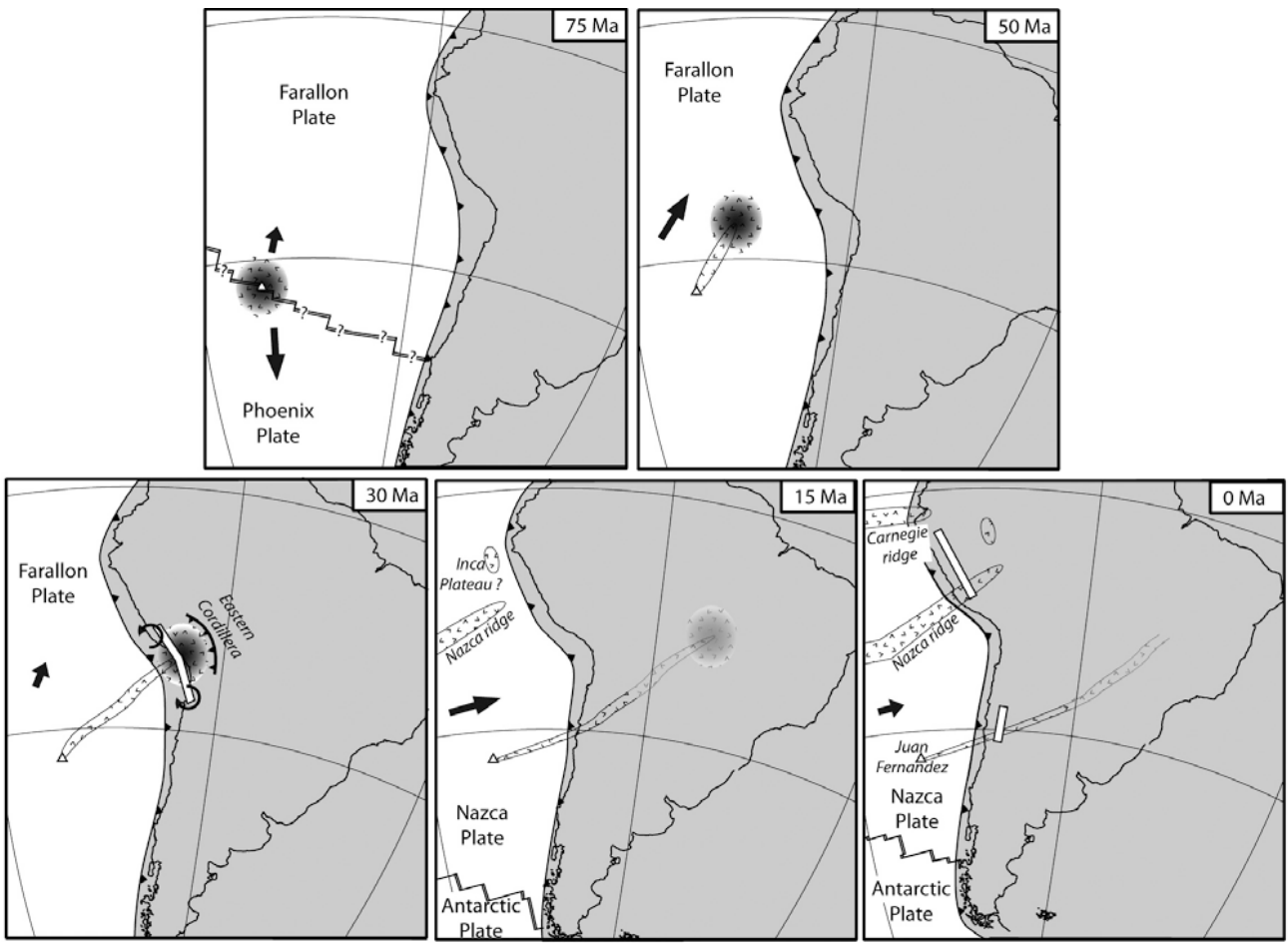

**Figure 4: Scenario of ridge subduction, slab geometry, and geological evolution of the Andes. White bars mark horizontal slab segments. ~75 Ma ago, the ridge separating the Phoenix and Farallon plate is moving southward. The Juan Fernandez hot spot is located close to the ridge. Question marks indicate that the position of the Phoenix-Farallon ridge is poorly constrained since the ridge has essentially been subducted afterwards. An oceanic plateau is forming above the Farallon plate, resulting from the hot spot activity. 50 Ma ago, this oceanic plateau is migrating northward above the plate. 30 Ma ago, the oceanic plateau subducts beneath the Central Andes, resulting in horizontal subduction beneath the Central Andes, tectonic rotations (Bolivian orocline) and shortening of the Eastern Cordillera. 15 Ma ago, the slab is inclined again beneath the Central Andes. The Juan Fernandez ridge is rapidly migrating southward. Present-day situation: the Juan Fernandez ridge is subducting for ~10 Ma close to 33°S, and a new horizontal slab segment has appeared beneath Chile. Another slab segment is present beneath Peru, resulting from the subduction of the Nazca ridge.**

### Impact of horizontal subductions on the overriding plate tectonic regime

High friction on the interplate contact generates earthquakes along horizontal segments of subduction. Many authors also noted that intraplate continental seismicity is more abundant above flat-slab regions than over the neighboring steep slab segments (Jordan et al., 1983; Smalley et al., 1993; Gutscher et al., 2000). Gutscher et al. (2000) quantified the seismic energy released within the overriding plate through the Andean chain during the 20[th] century, and observed that it has been 3 to 5 times larger above horizontal slab segments than elsewhere. Although larger released seismic energy may be attributed to a different upper plate rheology above flat-slab segments, it may also evidence the effects of the larger interplate contact area in horizontal slab segments that further shortens the upper plate. Higher shortening rates in the overriding plate following horizontal subduction have been reproduced by Espurt et al. (2008) in analogue experiments. The more the wedge narrows during slab flattening, the more interplate stresses resist further flattening and plate convergence increases. The major effect of horizontal subduction is to make shortening in the overriding plate prograde away from the trench, above the extremity of the flat slab segment. This phenomenon is observed in the present-day strain regime of the continental plate (Jordan and Allmendinger, 1986; Pardo et al., 2002; Ramos et al., 2002; Siame et al., 2005). In northwestern Argentina for instance, the tectonic activity migrated eastward during slab flattening (Ramos et al., 2002). Continental shortening now concentrates along the Sierra de Pie de Palo, at the boundary between the Sierras Pampeanas and the Cordillera. Increasing shortening above horizontal slab segments should also result in differential shortening accommodated along the active margin orogen, in block rotations, and should affect the orientation of the trench.

This may have also occurred in the past in the Central Andes: the Late Eocene-Oligocene uplift of the Eastern Cordillera (e.g., Lamb and Hoke, 1997; McQuarrie et al., 2005) may have been triggered by the horizontal subduction located below Southern Peru-Northern Chile at that time (James and Sacks, 1999). The growth of the Eastern Cordillera in the Central Andes is associated to larger shortening of the overriding plate at that latitude than in the Andes at the latitude of Central Peru and Central Chile at the same epoch. Extra-shortening in the Central Andes is recorded by rotations in the fore-arc of Northern Chile and Southern Peru. Paleomagnetic data evidence counterclockwise rotations in the Southern Peru forearc (37° average rotations) and clockwise rotations in Northern Chile (Roperch et al., 2006; Arriagada et al., 2008). These authors show that most of these rotations occurred in the Late Eocene-Oligocene period (see above, Figure 4), as a consequence of differential shortening focused in the Eastern Cordillera. Flat slab promoting interplate friction, it may explain the increase of continental shortening in the Eastern Cordillera, despite the fact that Oligocene is a period of slow convergence between the subducting plate and the South American continent (Figure 1). At present-day, the Central Andes are mature and their elevation can no longer increase because buoyancy forces balance tectonic forces (e.g., Dalmayrac and Molnar, 1981; Husson and Ricard, 2004). This is evidenced by the observed stress regime that is on average neutral within the Altiplano, as revealed by the coexistence of extension, strike-slip, and compression (see World Stress Map, Heidbach et al., 2008). The locus of deformation is thus transferred eastward in the lowland Subandean Zone, despite the fact that the Nazca slab went back to a steeper dip underneath the Central Andes.

The dynamics of subduction beneath South America and the Andean orogeny shares common features with the subduction beneath North America and the Laramide orogeny. A major episode of horizontal subduction has been proposed to explain the stress transfer to the base of the lithosphere in the foreland of the Rocky Mountain (Dickinson and Snider, 1978). Bird (1998) analyses the possible causes of continental shortening, and finds that the surface of interplate contact is a better predictor of continental deformation than the convergence velocity between either the Farallon or the Kula plate with the continent, confirming that horizontal subduction promotes continental shortening.

To sum up, the two episodes of volcanic quiescence (present and mid-Eocene to Oligocene) seemingly mark the end of the *Incaic* and *Quechua* phases of increased compression in many regions of the Andes. The timing of the events therefore makes it natural to establish a causal relationship between fostered compression, mountain building, slab flattening, and volcanic quiescence.

**Impact of horizontal subduction on the convergence velocity**

The convergence velocity results from the dynamic interaction between the driving and resisting forces that excite both the subducting and overriding plates. To quantify the convergence velocity, one option is to evaluate independently the different contributors (e.g. Meade and Conrad, 2008). One may alternatively consider the system globally, wherein both plates are driven towards the trench by driving forces (slab pull, ridge push, mantle drag) and resisted by mantle drag, slab bending and interplate friction of equally unknown magnitudes. Note that the overriding plate plays a fundamental role in the process of subduction (e.g., Conrad and Lithgow-Bertelloni, 2002; Clark et al., 2008; Bonnardot et al., 2008; Guillaume et al., 2009; Yamato et al., 2009; Capitanio et al., 2010), and particularly in South America (e.g., Silver et al., 1998). Making an accurate forward estimate of each contributor is tricky because the associated uncertainties are likely above the variations of their equilibrium that change the dynamics at large scale.

Here we focus on the changes in interplate force through time that modulate the convergence rate by resisting plate motion at variable rates (e.g. Norabuena et al., 1999). In particular in the Andes, the shear force has changed due to the variations in the geometry of the slab, from flat to steep angles of subduction. In their model, Meade and Conrad (2008) consider that this force is restricted to the coupling of plates in the brittle unit. It is probably more appropriate to consider that coupling applies all along some sort of subduction channel (e.g., Cloos and Shreve, 1988; Yañez et al., 2004; Gerbault et al., 2009) that is sheared between the plates (Figure 5). Indeed, the shift of overriding plate shortening towards the interior of the continent (see above) can be viewed as straightforward evidence that interplate friction transfers horizontal compression within the upper plate along the entire horizontal subduction contact area. Assuming all is linear and describing the deformation with a Newtonian rheology, the shear force $F_s$ may then write $F_s \sim \eta_c UL/D$, where $\eta_c$, $D$ and $L$ are the effective viscosity, thickness and length of the subduction channel between both plates; $U$ is the shear rate, i.e. the convergence rate. If the only parameter that changes in the force balance on the interplate zone is $L$, plate velocity responds proportionally. In other words, for a given force $F_s$, if the length of the subduction channel increases by a factor of 2 (approximately the ratio between the lengths of flat slab channels to steep slabs channels), $U$ will decrease by a factor of 2. Note that the choice of a viscous rheology is not relevant to our purpose as all we need to keep in mind is that $U$ is a decreasing function of $L$.

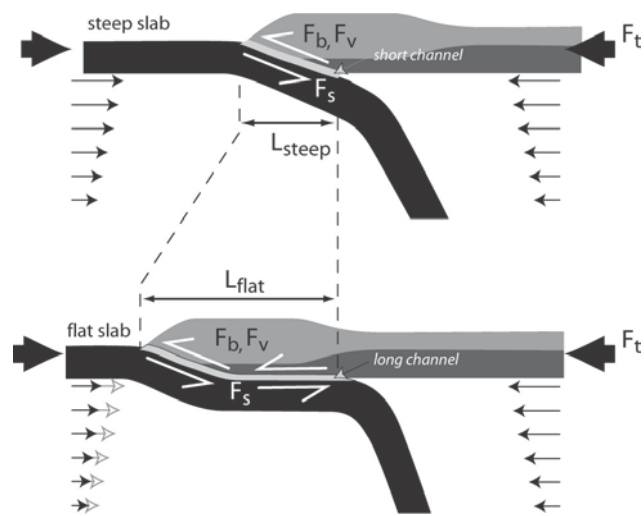

**Figure 5: Main forces controlling the dynamics of subduction below the Andes. Horizontal subduction increases the length of the subduction channel (L), which in turn may decrease the convergence rate. Fb: Buoyancy force resulting from the load of the belt; Fv : Force spent to shorten the continental plate; Fs : Shear force applied by the subducting plate on the overriding continent; Ft : far-field "tectonic" forces.**

In the framework of this approximation applied to the entire convergent system, it is therefore the total surface of the interplate contact that controls plate convergence for a given force $F_s$. In the Nazca – South America system, that surface can be estimated from the Benioff zone at present-day, and from the geological records mentioned above for the past. For simplicity we quantify the dimensions of this surface based on the present-day distribution of the segments of the Andean slab (see Figure 6 and Table 1). At the time of generalized slab steepness at ~20 Ma, the surface of the channel was lower by 20% to 65%. Given the above relationship and assuming all remained constant but the surface of the interplate contact, this estimate is in the right range to explain the velocity drop by ~40% (Figure 1). Similarly, the drop in convergence rates during Eocene could be associated with similar changes in the surface of the interplate contact. It is more hazardous to quantify the surface increase at that time for it is only based on geological inferences from the volcanism and distribution of the deformation. However, assuming that the 1100 km wide Bolivian segment was flat and that the lengths of coupling were comparable to that of present-day (~250±75 km for steep slabs and 700±75 km for shallow dipping slabs), the areal change of the interplate contact zone is ~35%, which is also in the correct range.

**Figure 6: Trench-perpendicular length of the interplate contact 20 Ma ago (dashed line) and at present-day (solid line). The surface of the interplate contact between Nazca and South America (grey area) increased in the Pliocene with the appearance of two major flat slab segments (see Table 1).**

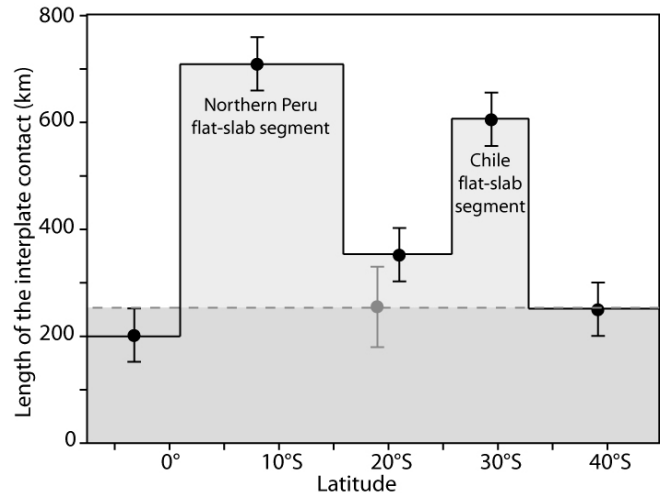

| Segment | Length (km) | L (km), 0 Ma | Surface ($10^3$ km$^2$), 0 Ma | L (km), 20 Ma | Surface ($10^3$ km$^2$), 20 Ma |
|---|---|---|---|---|---|
| **Ecuador (7.5°N-0.5°S)** | 875 | 200±50 | 175±44 | 250±75 | 219±66 |
| **Peru (0.5°S-16°S)** | 1700 | 700±50 | 1190±85 | 250±75 | 425±127 |
| **Bolivia (16°S-26°S)** | 1100 | 350±50 | 385±55 | 250±75 | 275±83 |
| **Central Chile (26°S-33°S)** | 765 | 600±50 | 459±38 | 250±75 | 191±57 |
| **South Chile (33°S-45°S)** | 1300 | 250±50 | 325±65 | 250±75 | 325±98 |
| **Total** | 5740 | | 2534±287 | | 1435±431 |

**Table 1: Geometrical properties of the subduction interface beneath the Andes at 0 Ma and 20 Ma.**

The geodynamic setting of the Nazca – South America system has not changed much since the rupture of the Farallon plate, 23 Ma ago. Thus, there is no *a priori* reason for the residual force applied to the subduction zone to have significantly changed since the closure of the Tethys and subsequent fostered westward motion of South America. However a few facts besides the above described interplate coupling are worth mentioning. First, slab pull may have decreased due to a decrease in the mean age of the Nazca subducting plate (see reconstructions by Müller et al., 2008).This situation was even more dramatic in North America in the Cenozoic: although interplate coupling decreased coevally to the gradual disappearance of the subducting Farallon plate (Sigloch et al., 2008), convergence rates decreased for the last 30 Ma while our model predicts that it should have increased. The fact that the mean subduction age underneath North America is extremely low (<10 Ma, see reconstructions by Müller et al., 2008) likely reveals that our model breaks down when the age-buoyancy of the subducting plate is too low. Similarly, the subduction of passive ridges may have further decreased the average density of the subducting plate, which may also contribute to the diminution of the subduction velocity.

Second, our estimate only holds if friction is the main break to convergence. Although Conrad and Hager (1999) numerical models predict that for old subducting plates, plate bending dissipates most of the slab's potential energy, they estimate that in young oceanic plate subduction zones, forces acting at the interplate contact together with dissipation in the mantle may absorb most of the energy released. In fact, plate bending dissipates a large amount of energy for subduction models wherein the radius of curvature is imposed and the slab is very stiff. If conversely the radius of curvature is set free, only slabs that are 3 to 4 orders of magnitude more viscous than the surrounding mantle do affect the subduction dynamics (Royden and Husson, 2006). In such case its overall effect is probably small in real Earth (e.g., Capitanio et al., 2009), for slab viscosity may be only 2 orders of magnitude higher than that of the mantle (e.g., Moresi and Gurnis, 1996; Funiciello et al., 2008; Loiselet et al., 2009). In particular for young plates such as Nazca and Farallon in the Cenozoic, the slab to mantle viscosity ratio may be low.

Third, the increasing load of the Andes has previously been perceived as a modifier of the coupling between the two plates, in which case its effect would become more and more important through time (Iaffaldano et al., 2006; Meade and Conrad, 2008). But the orogenic buoyancy force (sometimes referred to as gravitational potential energy) is not equal to the force at the interplate contact $F_s$. Instead, that force can be decomposed into the buoyancy force $F_b$ and the viscous force $F_v$ that corresponds to the force spent to deform the lithosphere. It may be conceived more clearly by looking at the early orogenic times, when $F_b$ was low. Neglecting $F_v$ implies that $F_s$ was also low, which essentially makes it impossible to develop a mountain belt, for otherwise causes and consequences are reversed. It is therefore more valid to state that $F_s = F_b + F_v$ and that the force $F_s$ that is responsible for mountain building is balanced by buoyancy and viscous forces, with $F_b$ gradually replacing $F_v$ during mountain building until a dynamic equilibrium is reached (Husson and Ricard, 2004), while the sum remains possibly unchanged. The equality can be further developed: those forces ultimately result from the balance between the driving far-field, "tectonic" forces $F_t$, eventually exciting the subduction zone with a residual force $F_s$. If the Andes developed in a geodynamic environment that remained steady for the last ~23 Ma, then both $F_t$ and $F_s$ remained steady. Going back to the stress balance above the subduction zone, it indicates that the effect of Andean building as described by its buoyancy force only shall not affect plate convergence for the total force $F_s$ does not depend on $F_b$ only but on the sum of the buoyancy and viscous forces. In our case study, it is assumed that the only parameter that has significantly changed is the coupling between the plates. That coupling variably resists the constant $F_s$ (or $F_t$) through time and the adjustment can be seen in the convergence rates.

Then, the decreasing eastward velocity of the Nazca plate may not result from the uplift of the Central Andes (Iaffaldano et al., 2006), but instead would mark the development of horizontal slab segments both below Central Peru and Central Chile-Argentina that increase the interplate coupling. This modification in the subduction dynamics is also accommodated partially by a decrease in the spreading rates in the South Atlantic (see fig. 2c of Conrad and Lithgow-Bertelloni, 2007), because the spreading force is opposed by the higher interplate shear force, and partially by a transmission of the deformation towards the Pacific that becomes sheared westward (Husson et al., 2008).

**Conclusion**

The decreasing trenchward velocity of the Nazca plate from the Late Miocene has been interpreted as a consequence of the rapid uplift of Central Andes either following lithosphere delamination (e.g. Garzione et al., 2006) or crustal thickening (Meade and Conrad, 2008; Iaffaldano et al., 2006). This period also corresponds however to the little discussed appearance of the two major present-day horizontal slab segments below South America. Slab flattening is thus also a candidate to explain the decreasing velocity of the Nazca plate, because it increases the interplate contact, which more efficiently resists sliding between the two plates (Espurt et al., 2008). Moreover, slab flattening below South America results from the subduction of buoyant ridges. In the paradigm where slab pull is regarded as the prominent driver of subduction zones (e.g. Forsyth and Uyeda, 1975; Chapple and Tullis, 1977; Funiciello et al., 2003), the subduction of buoyant slab units more efficiently resists slab pull and may also have decreased the subduction velocity.

We also propose that horizontal slab segments below South America significantly affected the dynamics of the Andes and plate kinematics in the Paleogene. Several pieces of evidence suggest that a large horizontal slab segment subducted below the Central Andes in the Eocene-Oligocene (James and Sacks, 1999). The subduction of an Upper Cretaceous oceanic plateau originated from the Juan Fernandez hot spot may have caused the appearance of this horizontal slab segment (Figure 4). Horizontal subduction may in turn have triggered the onset of shortening within the continental plate far from the trench and the growth of the Eastern Cordillera in the Central Andes, as well as the bending of the Bolivian orocline at that time (Roperch et al., 1996). Flat-slab subduction likely significantly increased the interplate contact along the subduction zone, which may explain the slower convergence velocity between the Farallon and South American plates. Then, the geometry of the subducting Farallon-Nazca plate below South America would have exerted a major control on the convergence velocity at trench.

Although beyond the scope of this paper, it would be interesting to study how horizontal subductions outside South America influenced both the overriding plate tectonic regime and plate kinematics. The best known example is the Laramide orogeny in North America, during which the horizontal subduction promoted stress transfer in the foreland of the Rocky Mountains. Subduction velocity beneath North America decreased despite the end of the flat-slab episode (Sigloch et al., 2008), probably because the subducting oceanic plate has become too young to be efficiently pulled by the weight of the slab.

Flat-slab subductions underneath South America result from the forced subduction of buoyant slab segments beneath the advancing continent. The westward motion of South America seems to be controlled by joint effects of the opening of the Atlantic Ocean and the closure of the Tethys Ocean (Silver et al., 1998). The oceanic plate entering the trench being progressively younger, oceanic plateaus more easily shift the plate to positively buoyant. Thus, horizontal slab segments also dramatically mark the arrival at the trench of a younger ocean. Horizontal subduction increases interplate friction and favors continental shortening, which in turn uplifts the Cordillera. Andean elevation results from complex processes, possibly including delamination of dense parts of the thickened lithosphere (Garzione et al., 2006), which explains why crustal shortening and uplift are not necessarily contemporaneous. We thus propose that flat-slab subduction is a major phenomenon that ultimately favors Andean growth and uplift.

*__Acknowledgments :__* This work has been supported by INSU "Reliefs de la Terre" and "DyETI" programs. Authors thank Federico Davila and an anonymous reviewer for their comments that greatly improved this manuscript.